\documentclass[aps,article,author-year,notitlepage,showpacs]{revtex4-1}
\usepackage{graphicx}
\usepackage{float}
\usepackage[T1]{fontenc}
\usepackage[latin1]{inputenc}
\usepackage{amssymb}
\usepackage{amsmath}
\include{epsf}
\epsfverbosetrue

\newcommand{\bea}{\begin{eqnarray}}
\newcommand{\eea}{\end{eqnarray}}

\makeatother
\begin{document}

\title{Renormalization group flow equations connected to the $n$PI effective action}

\author{M.E. Carrington}
\email[]{carrington@brandonu.ca} \affiliation{Department of Physics, Brandon University, Brandon, Manitoba, R7A 6A9 Canada}\affiliation{Winnipeg Institute for Theoretical Physics, Winnipeg, Manitoba}

\date{\today}

\begin{abstract}
In this paper we derive a hierarchy of integral equations from the 4PI effective action which have the form of Bethe-Salpeter equations. We show that the vertex functions defined by these equations can be used to truncate the exact renormalization group flow equations. This truncation has the property that the flow is a total derivative with respect to the flow parameter. We also show that the truncation is equivalent to solving the $n$PI equations of motion. This result establishes a direct connection between two non-perturbative methods.  
\end{abstract}


\large
\maketitle

\section{Introduction}
\vspace{5pt}

There is much interest in the study of non-perturbative systems, which cannot be solved by exploiting the existence of a small expansion parameter. In this paper we discuss two formalisms that have been proposed to address non-perturbative problems: $n$-particle irreducible ($n$PI) effective theories \cite{Jackiw1974,Norton1975}, and  
the exact renormalization group (RG) \cite{Wetterich1993,Ellwanger1993,Tetradis1994,Morris1994}.  
The $n$PI formalism has been used to study finite temperature systems (see for example \cite{Blaizot1999,Berges2005a}),  
non-equilibrium dynamics and subsequent late-time thermalization (see \cite{Berges2001} and references therein), and transport coefficients \cite{Aarts2004,Carrington2008,Carrington2009,Carrington2010b}.  The exact RG has been applied to a variety of problems (for reviews see \cite{Morris-98,Bagnuls2001,Berges2002,Pawlowski2005,Delamotte2007,Rosten2007}). 

It has been proposed that the hierarchy of RG flow equations could be truncated at the level of the first equation using the Bethe-Salpeter (BS) equation derived from the 2PI effective action \cite{Blaizot2011,Blaizot2010}. 
The flow of the 2-point function is a total derivative with respect to the flow parameter, and the integral of the flow equation gives an integral equation whose solution is equivalent to the equation of motion (eom) for the 2-point function from the 2PI effective action. In this paper we show that the 4PI effective action produces two BS equations that can be used to truncate the RG flow equations at the level of the second equation, and that the resulting flow equations for the 2- and 4-point functions are total derivatives whose integrals give the 4PI eom's. This result is surprising.  Since the full hierarchy of RG flow equations are obtained using a single bi-local source term, one does not expect a connection to the nPI formalism beyond the lowest 2PI level. It suggests that a BS truncation at arbitrary orders produces equations whose integrals give the $n$PI eom's, and establishes a direct connection between two non-perturbative methods. 
It also means that the truncation of the RG equations at any level of the hierarchy can be systematically extended by adding more and more skeleton diagrams to the effective action. For the $n$PI formalism, there could be a practical advantage in reformulating the integral equations as flow equations, because initial value problems are usually easier to solve than non-linear integral equations. Furthermore, regarding the vertices from the $n$PI effective theory as flow equations gives new insight into the problem of how to renormalize the $n$PI effective theory for $n>2$ \cite{Carrington2012}.

The paper is organized as follows. In section \ref{section:notation} we define our notation. In sections \ref{section:RGflow} and \ref{section:nPI} we give brief reviews of the RG flow equations and $n$PI effective action. In section \ref{section:LOtruncation} we review the derivation of the BS equation for the 4-point vertex from the 2PI effective action, and the procedure to use this equation to truncate the lowest order RG flow equation. 
%In section \ref{section:HOtruncation-summary} we summarize the results we have obtained using a BS truncation at higher orders. 
In section \ref{section:HOBS} we calculate the higher order BS equations that we will need, and in section \ref{section:HOtruncations} we show how they can be used to truncate the renormalization group equations at higher orders. We present our conclusions in section \ref{section:conclusions}. Some details are left to the appendices. 

\section{Notation}
\label{section:notation}

We work with a scalar field theory with quartic coupling and consider only the symmetric case where the expectation value of the field is zero. 
We define all propagators and vertices with factors of $i$ so that figures look as simple as possible:
lines, and intersections of lines, correspond directly to propagators and vertices, with no additional factors of plus or minus $i$.
The classical action is:
\bea
\label{scl}
S[\varphi]=\int d^dx \bigg(\frac{1}{2}\varphi(x)^2-\frac{m^2}{2}\varphi(x)^2-\frac{i}{\;4!}\lambda \varphi(x)^4\bigg)\,,
\eea
and the bare propagator is defined:
\bea
\label{bare-prop-def}
G_0^{-1}(x,y) =-i\frac{\delta^2 S[\varphi]}{\delta\varphi(x)\delta\varphi(y)}\,.
\eea
We use a compactified notation in which the space-time coordinates are represented by a single numerical subscript. For example, 
the propagator in equation (\ref{bare-prop-def}) is written $G_{ij}:=G(x_i,x_j)$.  We also use an Einstein convention in which a repeated index implies an integration over space-time variables.
Using this notation we define the generating functionals:
 \bea
 \label{genFcn}
&& Z[J]=\int d\varphi {\rm Exp}[i(S+J_i \varphi_i)]\,,\nonumber\\
&& W[J]=-i{\rm Ln} Z[J] \,,\nonumber\\
&& \Gamma[\phi]=W[J] - J_i\frac{\delta W}{\delta J_i}\,.
 \eea
The functional $W[J]$ is the generator of connected functions which are defined:
\bea
\label{Wders}
V^c_{x_1,x_2,x_3, \cdots x_k} = \langle\varphi_{x_1}\varphi_{x_2}\varphi_{x_3}\dots\varphi_{x_k}\rangle_c = -(-i)^{k+1}\frac{ \delta^k W}{\delta J_{x_k}\dots \delta J_{x_3}\delta J_{x_2}\delta J_{x_1}}\,.
\eea
The functional $\Gamma[\phi]$ generates 1-line irreducible, or proper, $n$-point functions. Using the notation  $\Gamma = - i\Phi$ they are defined:
\bea
\label{properDefn}
V_{x_1,x_2,x_3, \cdots x_k} &&=\Phi^{(n)}(x_1,x_2,x_3,\cdots x_n)=  \frac{\delta^k \Phi[\phi]}{\delta \phi_{x_k}\dots \delta \phi_{x_3}\delta \phi_{x_2}\delta \phi_{x_1}}\,.\nonumber
\eea
Equations (\ref{bare-prop-def}), (\ref{genFcn}) and (\ref{properDefn}) give\footnote{The minus sign on the left side of this equation comes from the fact that the effective action is defined as a functional of the propagator instead of the inverse propagator.} \bea
\label{sign-flip}
-\Phi^{(2)}(x,y) = G^{-1}(x,y) = G_0^{-1}-\Sigma(x,y)\,.
\eea
These $n$-point functions are invariant under translations of the co-ordinates and therefore in momentum space they depend on $n-1$ momenta. We use  incoming momenta and the convention that the Fourier transformed $n$-point functions are defined without the 4-dimensional delta function that enforces the conservation of momentum:
\bea
\label{PHI-mom}
(2\pi)^d\delta^d(p_1+p_2+\cdots p_n)\Phi^{(n)}(p_1,p_2,\cdots p_n)  = \prod_{k=1}^n \int d^d x_k \;e^{i\sum_{j=1}^n p_j x_j}\; \Phi^{(n)}(x_i \cdots x_n)\,.
\eea
We often use the shorthand notation:
\bea
\label{shorthand}
\Phi^{(n)})(p_1,p_2,\cdots p_{n-1},-p_1-p_2-\cdots p_{n-1}) = \Phi^{(n)})(p_1,p_2,\cdots p_{n-1})\,,
\eea
 for example we write $\Phi^{(2)}(p,-p) = \Phi^{(2)}(p) = -G^{-1}(p)$.

\section{Renormalization group flow equations}
\label{section:RGflow}

In this section we give a brief summary of the RG flow equations (for reviews see \cite{Morris-98,Bagnuls2001,Berges2002,Pawlowski2005,Delamotte2007,Rosten2007}). The RG is constructed by building a family of theories indexed by a continuous parameter $\kappa$ with the dimension of a
momentum, such that fluctuations are smoothly taken into account as $\kappa$ is lowered from
the microscopic scale $\Lambda$ (at which the couplings are defined) down to zero. To accomplish this, we add to the original action a non-local term which is quadratic in the fields:
\bea
\Delta S_{\kappa}[\varphi] = \frac{1}{2}\int dQ\, {\cal R}_\kappa(q)\varphi(q)\varphi(-q) \,,~~~dQ:=\frac{d^dq}{(2\pi)^d}\,.
\eea
The function ${\cal R}_\kappa$(q) is chosen so that it approaches zero for $q\gtrsim \kappa$ and $\kappa^2$ for $q\ll \kappa$. The first of these properties ensures that modes $\varphi(q\gtrsim\kappa)$ are unaffected, and the second suppresses the contribution of the modes $\varphi(q\ll \kappa)$  by giving them a mass $\sim\kappa$.

\begin{figure}[H]
\begin{center}
\includegraphics[width=10cm]{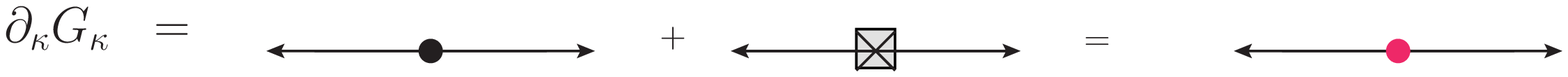}
\end{center}
\caption{Diagrammatic representation of equation (\ref{Gkappa-deriv}). The solid dot represents the insertion $\partial_\kappa R_\kappa$, the box with the cross is $\partial_\kappa\Phi_\kappa^{(2)}$, and the grey dot (red on-line) is the sum. Arrows on the ends of the lines indicate that propagator legs are attached. \label{fig2:Gkappa-deriv}  }
\end{figure}

Generating functionals are defined as in equation (\ref{genFcn}) with the action $S$ replaced by $S+\Delta S_\kappa$, so that each generating functional now depends on the flow parameter $\kappa$. Differentiating with respect to $\kappa$ we obtain:
 \bea
 \partial_\kappa W_\kappa =  \frac{1}{2}\int dQ \partial_\kappa {\cal R}_\kappa \langle\varphi(q)\varphi(-q)\rangle_\kappa = \frac{1}{2}\int dQ \partial_\kappa {\cal R}_\kappa  G_\kappa(q,-q)\,,
\eea
where the subscript on the expectation value indicates that it depends on $\kappa$ (we remind the reader that we are considering the symmetric theory for which $\langle \varphi\rangle_\kappa=0$). It is easy to obtain the corresponding expression for the effective action. 
Using $\Phi_\kappa = i\Gamma_\kappa$ and defining $R_\kappa = i {\cal R}_\kappa$ we obtain: 
 \bea
 \label{1stE}
\partial_\kappa \Phi_\kappa= \frac{1}{2}\int dQ \partial_\kappa R_\kappa \langle\varphi(q)\varphi(-q)\rangle_\kappa = \frac{1}{2}\int dQ \partial_\kappa R_\kappa  G_\kappa(q,-q)\,,
 \eea 
and equation (\ref{sign-flip}) becomes:
\bea
\label{Gkappa-def}
-G_\kappa^{-1} = R_\kappa + \Phi_\kappa^{(2)}\,,
\eea
which gives:
\bea
\label{Gkappa-deriv}
\partial_\kappa (G_{xy})_\kappa = (G_{xa})_\kappa\,\partial_\kappa(R_\kappa+\Phi_\kappa^{(2)})_{ab}\,(G_{by})_\kappa\,.
\eea
The diagrammatic notation we will use for equation (\ref{Gkappa-deriv}) is shown\footnote{Figures in this paper are drawn using jaxodraw \cite{jaxo}.} in figure \ref{fig2:Gkappa-deriv}. 
\par

\begin{figure}[H]
\begin{center}
\includegraphics[width=10cm]{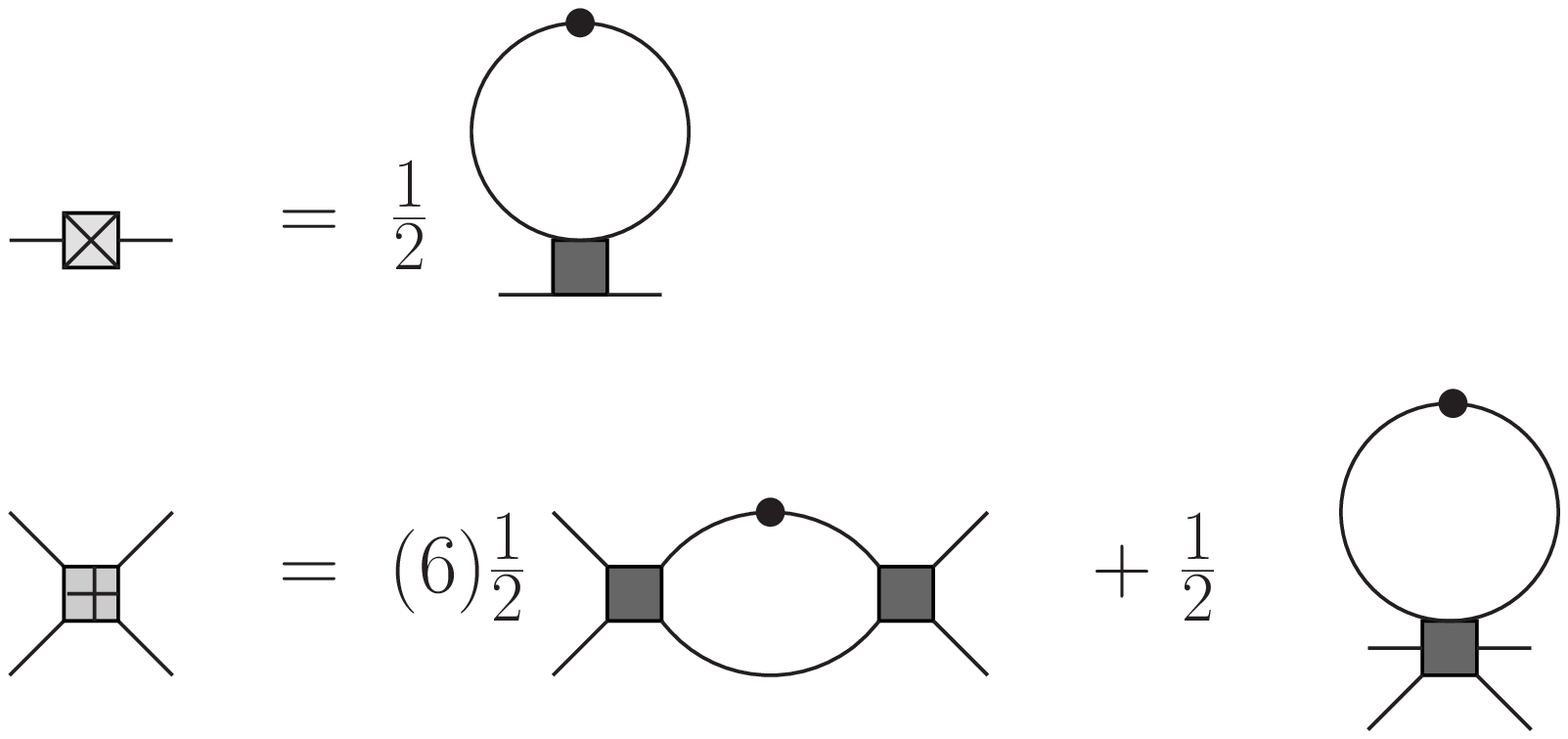}
\end{center}
\caption{Diagrammatic representation of equations (\ref{flow1}) and (\ref{flow2}). Dark grey boxes with 2, 4 and 6 legs represent $\Phi_\kappa^{(2)}$, $\Phi_\kappa^{(4)}$ and $\Phi_\kappa^{(6)}$, respectively. Boxes with crosses through them (on the left side of the figure) represent the derivative with respect to the flow parameter of the corresponding vertex. The solid dot on a propagator represents the insertion $\partial_\kappa R_\kappa$.  \label{fig:renormGroup}  }
\end{figure}

Functionally differentiating equation (\ref{1stE}) with respect to $\phi$ produces a hierarchy of equations known as the exact RG flow equations. The first two equations in this hierarchy are ($p_4=-p_1-p_2-p_3$):
\bea
\label{flow1}
\partial_\kappa \Phi_\kappa^{(2)}(p) = &&\frac{1}{2}\int dQ \,\partial_\kappa R_\kappa (q)G^2_\kappa(q) \Phi_\kappa^{(4)}(p,-p,q,-q)\,, \\
\partial_\kappa \Phi_\kappa^{(4)}(p_1,p_2,p_3,p_4) =&& \frac{1}{2}(6)\int dQ\, \Phi_\kappa^{(4)}(p_1,p_2,-q)G_\kappa(q) \partial_\kappa R_\kappa(q)G_\kappa(q) G_\kappa(p_1+p_2-q) 
\Phi_\kappa^{(4)}(q,p_3,p_4)\nonumber\\
\label{flow2}
&& + \frac{1}{2}\int dQ \,G_\kappa(q)\partial_\kappa R_\kappa(q)G_\kappa(q) \Phi_\kappa^{(6)}(q,p_1,p_2,p_3,p_4)\,.
\eea
These equations are shown in figure \ref{fig:renormGroup}. 
The factor (6) in equation (\ref{flow2}) and figure \ref{fig:renormGroup} is a short-hand notation which means that there are 6 permutations of external legs only one of which is explicitly indicated. These correspond to the 4! ways to permute the 4 external legs of the diagram, divided by a factor 2$\cdot$2=4 to account for the fact that the vertices $\Phi_\kappa^{(4)}(p_1,p_2,-q)$ and $\Phi_\kappa^{(4)}(q,p_3,p_4)$ are symmetric under permutation of their legs. The 5 terms that are not written can be produced from the one which is using the variable changes: $p_2 \leftrightarrow p_1$, $p_2 \leftrightarrow p_4$, $\{p_1,p_2\} \leftrightarrow \{p_2,p_3\}$, $\{p_1,p_2\} \leftrightarrow \{p_2,p_4\}$, $\{p_1,p_2\} \leftrightarrow \{p_3,p_4\}$. Throughout this paper we will use this notation: numerical factors in brackets in equations (figures) represent additional terms that correspond to permutations of external indices that are not written (drawn). 
\par

The RG flow equations form an infinite coupled hierarchy: the equation for $\Phi^{(2n)}$ involves $\Phi^{(2)}$ and $\Phi^{(2[n+1])}$. In order to do calculations, one must truncate the hierarchy. This is a common feature of non-perturbative methods, and often leads to difficulties (see for example \cite{Smit2003,Zaraket2004,Binosi2009,Serreau2010}). 
%In this paper we explore the relationship between the RG flow equations and the integral equations produced by the $n$PI formalism. 

\section{The $n$PI effective action}
\label{section:nPI}

\begin{figure}[H]
\begin{center}
\includegraphics[width=11cm]{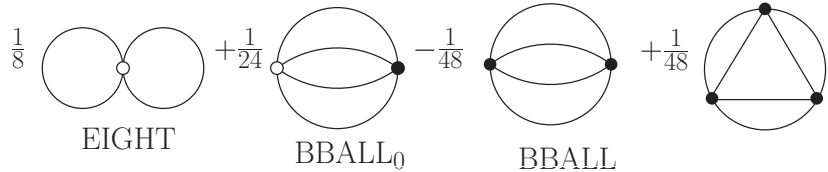}
\end{center}
\caption{The functional $\Phi_2$ for the 4-Loop 4PI effective action. Bare vertices are denoted by open circles and effective vertices are solid dots. \label{fig:PhiAandB-2}}
\end{figure}

The $n$PI effective action is obtained by taking the $n$th Legendre transform of the generating functional which is constructed by coupling the field to $n$ source terms:
\bea
\label{genericGamma}
&& Z[J,R,R^{(3)},R^{(4)},\dots]=\int d\varphi  \;{\rm Exp}[i\,{\cal X}]\,,\\[2mm] &&{\cal X}=S_{cl}[\varphi]+J_i\varphi_i+\frac{1}{2}R_{ij}\varphi_i\varphi_j + \frac{1}{3!}R^{(3)}_{ijk}\varphi_i\varphi_j\varphi_k + \frac{1}{4!} R^{(4)}_{ijkl}\varphi_i\varphi_j\varphi_k\varphi_l +\cdots\,,\nonumber\\[4mm]
&&W[J,R,R^{(3)},R^{(4)},\dots]=-i \,{\rm Ln} Z[J,R,R^{(3)},R^{(4)},\dots]\,,\nonumber\\[4mm]
&&\Gamma[\phi,G,U,V\dots] = W - J_i\frac{\delta W}{\delta J_i} - R_{ij}\frac{\delta W}{\delta R_{ij}} - R^{(3)}_{ijk}\frac{\delta  W }{\delta R^{(3)}_{ijk}} - R^{(4)}_{ijkl}\frac{\delta  W }{\delta R^{(4)}_{ijkl}}  -\cdots\nonumber
\eea
For future use we note the relations:
\bea\label{defcon}
\frac{\delta  W }{\delta J_i} &&= \langle\varphi_i\rangle = \phi_i \,,\\
2\frac{\delta  W }{\delta R_{ij}} &&= \langle\varphi_i \varphi_j\rangle = G_{ij}+\phi_i \phi_j \,.
\eea
The $n$PI effective action is obtained from the last line of Eq. (\ref{genericGamma}) and can be written:
\bea
\label{gammaGen}
\Gamma[\phi,G,U,V\dots] &&= -i \Phi[\phi,G,U,V\dots] \\
&&=S_{cl}[\phi]+
    \frac{i}{2} {\rm Tr} \,{\rm Ln}G^{-1}  +
\frac{i}{2} {\rm Tr}\left[ \left(G^0\right)^{-1} G\right] -i\Phi_2[\phi,G,U,V\dots]~~+~~{\rm const} \,.\nonumber
\eea
The term $\Phi_2[\phi,G,U,V\dots]$ contains all contributions to the effective action which have two or more loops.
For example, for the 4-Loop 4PI effective action \cite{Berges2004,Carrington2004,Carrington2010b} in the symmetric theory  $\Phi_2$  is shown in figure \ref{fig:PhiAandB-2}.
\par
\begin{figure}[H]
\begin{center}
\includegraphics[width=13cm]{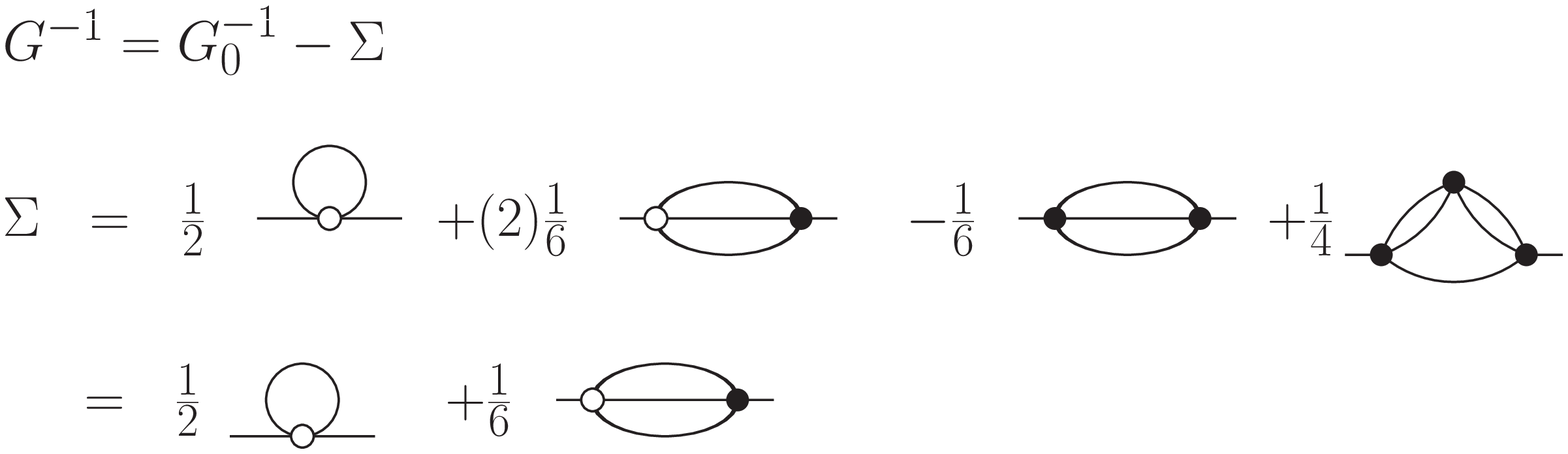}
\end{center}
\caption{Equation of motion for the self-energy from the 4-Loop 4PI effective action. The second line illustrates the notation we use throughout this paper in which different permutations of external indices are indicated with a bracketed numerical factor.  In the third line we have used the equation of motion for the 4-point function which is shown in figure \ref{fig:Veom}. \label{fig:SEall-betaS}}
\end{figure}
\begin{figure}[H]
\begin{center}
\includegraphics[width=15cm]{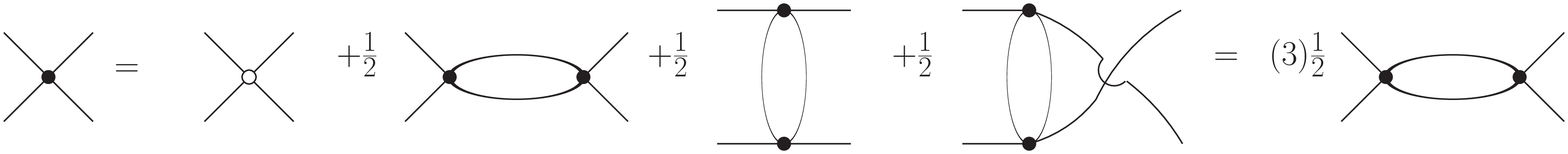}
\end{center}
\caption{Equation of motion for the 4-point function from the 4-Loop 4PI effective action. In the last section of the figure  different permutations of external indices are indicated with a bracketed numerical factor.  \label{fig:Veom}}
\end{figure}

The self consistent propagator and vertex are obtained through the variational principle by solving the equations produced by taking the functional derivative of the effective action and setting the result to zero. For the 4PI effective theory this gives:
\bea
\label{Geom}
&& -\frac{\delta \Phi}{\delta G_{xy}}  = - G^{-1}_{xy} + (G^{-1}_0)_{xy}-\Sigma_{xy}=0\,,~~~\Sigma_{xy}=2\frac{\delta\Phi_2}{\delta G_{xy}}\,, \\
\label{Veom}
&& 4! G^{-1}_{xx^\prime} G^{-1}_{xx^\prime} G^{-1}_{xx^\prime} \frac{\delta \Phi}{\delta V_{x^\prime y^\prime w^\prime z^\prime }} = -V_{xywz}+\lambda \delta_{xy}\delta_{xw}\delta_{xz}+4! G^{-1}_{xx^\prime} G^{-1}_{yy^\prime} G^{-1}_{ww^\prime} G^{-1}_{zz^\prime} \frac{\delta \hat\Phi_2}{\delta V_{x^\prime y^\prime w^\prime z^\prime} }   = 0\,.
\eea
The minus sign on the left side of equation (\ref{Geom}) is related to the minus sign in equation (\ref{sign-flip}) and is discussed in footnote 1.
The term $- G^{-1}_{xy}$ comes from the 1-loop terms in the effective action and is moved to the other side of the equation to produce the usual form of the Dyson equation. This is shown in figure \ref{fig:SEall-betaS}.  In equation (\ref{Veom}) the term $-V_{xywz}$ is produced by the basketball (BBALL) diagram in figure \ref{fig:PhiAandB-2} and is moved to the other side to produce the equation of motion shown in figure \ref{fig:Veom}. 
\par
\par

\section{Lowest order truncation}
\label{section:LOtruncation}

\subsection{Bethe-Salpeter equation from the 2PI effective action}
\label{section:standardBS}

It is well known that the 2PI effective action can be used to obtain a 4-point vertex called the Bethe-Salpeter vertex \cite{vanHees2002}. In this section we review the derivation of this equation.  We calculate the functional derivative of the effective action with respect to the 2-point function $G_{kl}$ and the source $R_{ij}$. 
We do the calculation in two different ways and equate the results. First, we use the last line in equation (\ref{genericGamma}) with the derivatives $\delta W/\delta J$ and  $\delta W/\delta R$ written in terms of the expectation value and propagator using equation (\ref{defcon}). Differentiating we obtain:
\bea
\label{preside1}
\frac{\delta}{\delta R_{ij}} \frac{\delta}{\delta G_{kl}}\Phi &&= i \frac{\delta}{\delta R_{ij}}\bigg(\frac{\delta J_x}{\delta G_{kl}}\frac{\delta W}{\delta J_x} +  \frac{\delta R_{xy}}{\delta G_{kl}}\frac{\delta W}{\delta R_{xy}} - \frac{\delta J_x}{\delta G_{kl}} \phi_x \nonumber\\[2mm]
&&-\frac{1}{2}\frac{\delta R_{xy}}{\delta G_{kl}}\big(G_{xy}+\phi_x\phi_y\big)-\frac{1}{4} R_{xy}(\delta_{xk}\delta_{yl}+\delta_{xl}\delta_{yk})\bigg)\,.
\eea
Using (\ref{defcon}) the first and third, and the second and fourth terms on the right side cancel identically, and we are left with:
\bea
\label{side1}
\frac{\delta}{\delta R_{ij}} \frac{\delta}{\delta G_{kl}}\Phi = -\frac{i}{4}\big(\delta_{ik}\delta_{jl}+\delta_{il}\delta_{jk}\big)\,.
\eea
Now we repeat the calculation using equation (\ref{gammaGen}) for the effective action. We obtain:
\bea
\label{side2}
\frac{\delta}{\delta R_{ij}} \frac{\delta}{\delta G_{kl}}\Phi = \frac{\delta \phi_x}{\delta R_{ij}}\frac{\delta^2\Phi}{\delta\phi_x\delta G_{kl}}  +  \frac{\delta G_{xy}}{\delta R_{ij}}\frac{\delta^2\Phi}{\delta G_{xy}\delta G_{kl}}\,.
\eea
Since we consider only the symmetric theory, we drop all terms that correspond to vertices with an odd number of legs, which means that only the second term on the right side survives. Using (\ref{gammaGen}) we write:
\bea
\label{LAM020}
4\frac{\delta^2\Phi}{\delta G_{xy}\delta G_{kl}} && =4\frac{\delta^2(\Phi-\Phi_2)}{\delta G_{xy}\delta G_{kl}}+4\frac{\delta^2\Phi_2}{\delta G_{xy}\delta G_{kl}}\,,\nonumber\\[2mm]
&& =: \Lambda^{disco}_{xykl}+ \Lambda_{xykl} = -(G^{-1}_{xk}G^{-1}_{yl}+G^{-1}_{xl}G^{-1}_{yk})+\Lambda_{xykl}\,.
\eea
The term $\Lambda^{disco}_{xykl}$ represents all disconnected contributions and comes from the 1-loop terms in the effective action, and $\Lambda_{xykl}$ contains all contributions from $\Phi_2$. 

The last step is to calculate the derivative of the propagator with respect to the source $R$.
We have:
\bea
\label{GderR2}
\frac{\delta G_{xy}}{\delta R_{ij}} &&= \frac{\delta}{\delta R_{ij}}\big(\langle \varphi_x\varphi_y\rangle - \langle\varphi_x\rangle\langle\varphi_y\rangle\big)\,, \nonumber\\
&& = \frac{i}{2}\big(\langle \varphi_x \varphi_y \varphi_i \varphi_j\rangle - \langle \varphi_x\varphi_y\rangle  \langle \varphi_i\varphi_j\rangle  ~ + ~ \cdots\big)\,,  \nonumber\\
&& = \frac{i}{2}\big(G_{ia}G_{jb}G_{xc}G_{yd}V_{abcd}+G_{ix}G_{jy} +G_{iy}G_{jx}~+~\cdots \big)\,,
\eea
where the dots indicate expectation values which contain an odd number of field operators and are dropped since we are considering the symmetric theory. 

\begin{figure}[H]
\begin{center}
\includegraphics[width=10cm]{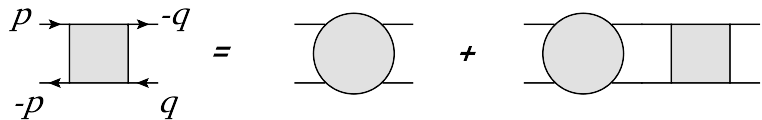}
\end{center}
\caption{Diagrammatic representation of the BS equation in equation (\ref{BSfirst-mom}). Boxes and circles represent the vertex $V$ and kernel $\Lambda$, respectively. \label{fig:standardBS}}
\end{figure}

We substitute equations (\ref{LAM020}) and (\ref{GderR2}) into (\ref{side2}) and set the result equal to the expression obtained in (\ref{side1}). This procedure gives:
\bea
0=\frac{i}{4} G_{ix} G_{jy}\,\big(-V_{klxy}+\Lambda_{klxy}+\frac{1}{2} \Lambda_{klab}G_{ac}G_{bd} V_{cdxy}\big)\,,
\eea
where we have used the fact that the vertex $V_{klxy}$ is symmetric with respect to permutations of any pair of indices, and the vertex $\Lambda_{klxy}$ is symmetric with respect to permutations of the first two indices, or the second two indices, or the interchange of the first pair and the second pair: $\Lambda_{klxy}=\Lambda_{lkxy}=\Lambda_{klyx}=\Lambda_{xykl}$. Truncating the external legs we obtain the standard form of the BS equation:
\bea
\label{BSfirst-coord}
V_{xykl}=\Lambda_{xykl}+\frac{1}{2} \Lambda_{xyab}G_{ac}G_{bd} V_{cdkl}\,.
\eea
We consider systems in thermal equilibrium for which the system is invariant under space-time translations. In this case equation (\ref{BSfirst-coord}) can be written in momentum space as:
\bea
\label{BSfirst-mom}
V(p,-p,q,-q) = \Lambda(p,-p,q,-q)+\frac{1}{2}\int dK\,\Lambda(p,-p,k,-k)G^2(k)V(k,-k,q,-q)\,.
\eea
Due to the translation invariance of the propagator, the 4-point function does not have general momentum arguments, but rather is restricted to the particular configuration indicated in equation (\ref{BSfirst-mom}). We will refer to these momentum arguments as ``diagonal.''
Equation (\ref{BSfirst-mom}) is shown diagrammatically in figure \ref{fig:standardBS}. 
\par

\subsection{Truncation of the lowest order renormalization group equation}
\label{section:LO-RGtruncation}

It has been proposed that the hierarchy of RG flow equations could be truncated using the BS equation derived in the previous section \cite{Blaizot2011,Blaizot2010}. The procedure is as follows. We extend equation (\ref{BSfirst-mom}) to the deformed theory by writing:
\bea
\label{BSfirst-mom-ext}
V_\kappa(p,-p,q,-q) = \Lambda_\kappa(p,-p,q,-q)+ \frac{1}{2}\int dK\,\Lambda_\kappa(p,-p,k,-k)G_\kappa^2(k)V_\kappa(k,-k,q,-q)\,.
\eea
In this equation the subscript on $\Lambda_\kappa$ indicates that the functional derivative which defines the kernel is taken with respect to the propagator, and the kernel is then evaluated at $G=G_\kappa$ (as defined in equation (\ref{Gkappa-def})). If we use $\Phi^{(4)}_\kappa = V_\kappa$ the equations (\ref{flow1}) and (\ref{BSfirst-mom-ext}) form a closed set. 
If the full effective action is used, the solution of this coupled system of equations  gives the full 2-point function, and the full 4-point function for diagonal momenta. This observation points out an important feature of the truncation: it can be systematically extended by adding more and more skeleton diagrams to the effective action. 

\begin{figure}[H]
\begin{center}
\includegraphics[width=10cm]{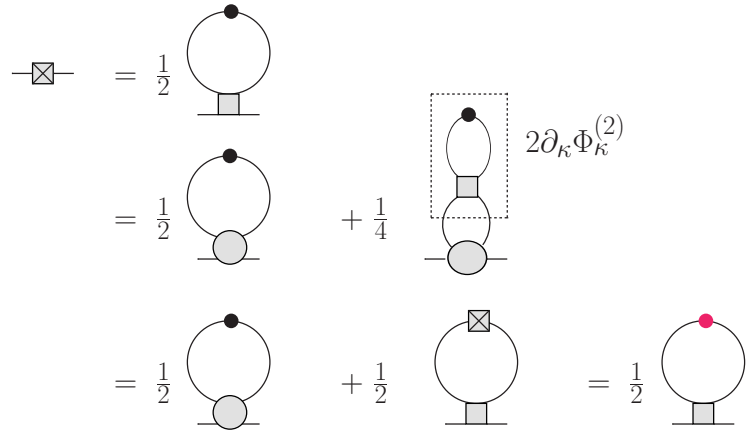}
\end{center}
\caption{The result obtained using the BS equation from section \ref{section:standardBS} to truncate the first RG flow equation in (\ref{flow1}).  \label{fig:JPsub}}
\end{figure}

The result of the truncation is easiest to see diagrammatically and is shown in figure \ref{fig:JPsub}. In the first line of this figure we replace the dark grey box in the tadpole diagram in figure \ref{fig:renormGroup} (which represents $\Phi_\kappa^{(4)}$) by the light grey box on the left side of figure \ref{fig:standardBS} (which represents $V_\kappa$). Using right side of figure \ref{fig:standardBS} we obtain the second line of figure \ref{fig:JPsub}. 
The box of dotted lines is just the insertion $2\partial_\kappa\Phi_\kappa^{(2)}$, using the first line of the figure. Inserting the first line into the second we obtain the first part of the third line. In the second part of the third line we use the notation in figure \ref{fig2:Gkappa-deriv} to represent the sum of terms $\partial_\kappa R_\kappa+\partial_\kappa \Phi^{(2)}$ by a small grey dot (red on-line).  
\par

An interesting aspect of this truncation is that it has the property that the flow is a total derivative with respect to the flow parameter $\kappa$. To prove this, we consider the 2-point function obtained from the 2PI effective action. Using equation (\ref{Geom}) and evaluating at $G=G_{\kappa}$ after taking the functional derivative we obtain:
\bea
\label{sigma-2PI-a}
(\Sigma_{ij})_\kappa = 2\frac{\delta\Phi_2}{\delta G_{ij}}\bigg|_{G_\kappa}=\Sigma_{ij}[G_\kappa]\,,
\eea
and therefore:
\bea
\label{sigma-2PI-b}
\partial_\kappa(\Sigma_{ij})_\kappa &&= 2\partial_\kappa (G_{kl})_\kappa \frac{\delta^2\Phi_2}{\delta G_{kl}\delta G_{ij}}\bigg|_{G_\kappa} = \frac{1}{2}\partial_\kappa (G_{kl})_\kappa (\Lambda_{ijkl})_\kappa\,, \nonumber\\
&& = \frac{1}{2}\big[(G_{ka})_\kappa \partial_\kappa (R_{ab}+\Phi^{(2)}_{ab}) (G_{bj})_{\kappa}\big] (\Lambda_{ijkl})_\kappa\,,
\eea
where we used equations (\ref{LAM020}) and (\ref{Gkappa-deriv}) in the first and second lines.
Equation (\ref{sigma-2PI-b}) is exactly the result for $\partial_\kappa\Phi_\kappa^{(2)}$ that is shown in the last line of figure \ref{fig:JPsub} and therefore we have obtained:
\bea
\label{2point-kappa}
\partial_\kappa \Phi_\kappa^{(2)} = \partial_\kappa \Sigma_\kappa\,.
\eea
Thus we have shown that the flow is a total derivative with respect to the flow parameter, and the integral of the flow equation gives an integral equation whose solution is equivalent to the equation of motion for the 2-point function from the 2PI effective action.

\section{Higher order BS equations}
\label{section:HOBS}

In this section we derive some higher order Bethe-Salpeter type equations from the 2PI and 4PI equations of motion. In the next section we will show how to use these equations to truncate the RG hierarchy at higher orders. Throughout this paper we use  circles to denote kernels and boxes are vertices obtained by solving integral equations. 

\subsection{Kernel notation}

We introduce some notation for the various kernels that we will encounter:
\bea
\label{LAM0nl}
\Lambda^{\rm disco}_{abcd\cdots rstuvxyz\cdots}+\Lambda_{abcd\cdots rstuvxyz\cdots} = 2^{\# G} 4!^{\# V} \big(G^{-1}_{rr^\prime}\cdots G^{-1}_{zz^\prime}\big)\frac{\delta^{(\# G)+(\# V)}\Phi}{\delta G_{ab}\delta G_{cd}\cdots\delta V_{r^\prime s^\prime t^\prime u^\prime}\delta V_{v^\prime x^\prime y^\prime z^\prime}\cdots}
\eea
The factors ${\# G}$ and ${\# V}$ indicate the number of $G$'s and $V$'s with respect to which the functional derivatives are taken.  
The inverse propagators truncate the legs that are left behind by the functional derivatives with respect to $V$ (there are $4\cdot ({\# V})$ inverse propagators in total) which produces an amputated kernel.
The definition of the kernel $\Lambda_{ab\cdots}$ excludes the disconnected contribution because this piece will always cancel in BS equations. For $\#G=2$ and $\#V=0$ equation (\ref{LAM0nl}) reduces to (\ref{LAM020}). 

Note that the kernels defined in (\ref{LAM0nl}) for the cases where only one derivative is taken are not really kernels, since the right side is just the equation of motion for the corresponding vertex. In this case we obtain an integral equation by moving the vertex to the other side of the equation, as explained under equation (\ref{Veom}).

Above equation (\ref{BSfirst-coord}) we commented on the symmetries of the 4-point vertex $\Lambda_{abcd}$ with respect to interchange of leg indices. In general, the vertices in equation (\ref{LAM0nl}) are symmetric with respect to the interchange of any two co-ordinates which came from the same $G$ or $V$ in the functional derivative. In addition, if there is more than one $G$ or $V$, one can interchange the full set of corresponding indices (in any order). For example, consider the 8-point kernel:
\bea
\label{LAMex}
\Lambda^{\rm disco}_{abcdrstu}+\Lambda_{abcdrstu} = 2^2 4! \big(G^{-1}_{rr^\prime}G^{-1}_{ss^\prime}G^{-1}_{tt^\prime}G^{-1}_{uu^\prime} \big)\frac{\delta^{3}\Phi}{\delta G_{ab}\delta G_{cd}\delta V_{r^\prime s^\prime t^\prime u^\prime}}\,.
\eea
Some permutations of the variables $\{a,b,c,d,r,s,t,u\}$ that produce the same vertex are: $a\leftrightarrow b$, $r\leftrightarrow s$ and $\{a,b\}\leftrightarrow \{c,d\}$. Two permutations that do not are $a\leftrightarrow c$ and $a\leftrightarrow r$.\footnote{Some authors insert a semi-colon between the two pairs of indices in the vertex $\Lambda_{abcd}$ to remind the reader that permutations of individual indices across the semi-colon are illegal. Using this notation we would write $\Lambda_{ab;cd;rstu}$ instead of $\Lambda_{abcdrstu}$. To keep equations as short as possible, we do not use the semi-colon and ask the reader to remember which sets of indices came from which functional derivative.}

In equilibrium, translation invariance means that the kernels will depend only on the differences of the co-ordinate indices of each element of the functional derivative. For example, the vertex in (\ref{LAMex}) does not depend on $x_a$ and $x_b$ individually, but only on the difference $x_a-x_b$. Similarly, it does not depend on $x_r$, $x_s$, $x_t$ and $x_u$ but only on (for example) $x_r-x_u$, $x_s-x_u$ and $x_t-x_u$. The consequence of this invariance is that in momentum space the kernel does not have a number of independent momentum arguments equal to the number of its legs. 
For example, the Fourier transform of the vertex in equation (\ref{LAMex}) has the form $\Lambda(p,-p,k,-k,q_1,q_2,q_3,-q_1-q_2-q_3)$ (using the shorthand notation introduced in equation (\ref{shorthand}) we sometimes write  $\Lambda(p,k,q_1,q_2,q_3)$). 
Using $\# G=2$ and $\# G=3$ we obtain 4- and 6-point kernels which have diagonal momentum arguments of the form $\Lambda(p,-p,q,-q)$ and $\Lambda(p,-p,q,-q,k,-k)$. Using $\# G=1$ and $\#V=1$ produces a 6-point kernel with momentum arguments $\Lambda(p,-p,q_1,q_2,q_3,-q_1-q_2-q_3)$ which we will call partially-diagonal. Thus the general 6-point function depends on 5 independent momenta, and the partially-diagonal and diagonal 6-point functions depend on 4 and 3 independent momenta, respectively.
Note that we always use the convention that the delta functions that enforce the conservation of momentum are removed from the Fourier transformed vertex (see equation (\ref{PHI-mom})).

\subsection{BS equation for a 6-point vertex from the 2PI effective action}
\label{section:6diag}

One can obtain a BS equation for a 6-point vertex from the 2PI effective action using the method that was used in section \ref{section:standardBS} for the 4-point function. To obtain a 6-point vertex we calculate the functional derivative $\delta^3 \Phi/\delta R\delta R\delta G$. Using the chain rule we have:
\bea
\label{BS222-a}
0=\frac{\delta^3 \Phi}{\delta R_{ab} \delta R_{cd} \delta G_{ef}}  = 
\frac{\delta^2 G_{xy}}{\delta R_{ab} \delta R_{cd}}\;\frac{\delta^2\Phi}{\delta G_{xy}\delta G_{ef}} + \frac{\delta G_{xy}}{\delta R_{cd}} \;\frac{\delta G_{rs}}{\delta R_{ab}}\; \frac{\delta^3\Phi}{\delta G_{rs}\delta G_{xy}\delta G_{ef}}\,,
\eea
where the zero on the left side is obtained from equation (\ref{side1}).
Using equation (\ref{LAM0nl}) the second derivative of the effective action with respect to the propagator gives the kernel $\frac{1}{4} (\Lambda^{\rm disco}_{xyef}+\Lambda_{xyef})$  and the third derivative  gives:
\bea
\label{LAM030}
8\frac{\delta^3\Phi}{\delta G_{rs}\delta G_{xy}\delta G_{ef}} && = \Lambda^{\rm disco}_{rsxyef}+ \Lambda_{rsxyef}\,.
\eea 
The disconnected contribution comes from the functional derivatives acting on the 1-loop piece of the effective action and is:
\bea
\Lambda^{\rm disco}_{rsxyef} && = -(G^{-1}_{e y} G^{-1}_{f s} G^{-1}_{r x} + G^{-1}_{e y} G^{-1}_{f r} G^{-1}_{s x}+G^{-1}_{e s}
   G^{-1}_{f y} G^{-1}_{r x}+G^{-1}_{e r} G^{-1}_{f y} G^{-1}_{s x} \nonumber\\[2mm]
   && ~~~+G^{-1}_{e x} G^{-1}_{f s}
   G^{-1}_{r y}+G^{-1}_{e s} G^{-1}_{f x} G^{-1}_{r y}+G^{-1}_{e x} G^{-1}_{f r} G^{-1}_{s
   y}+G^{-1}_{e r} G^{-1}_{f x} G^{-1}_{s y})\nonumber\\[2mm]
&& = (8)G^{-1}_{e y} G^{-1}_{f s} G^{-1}_{r x} \,.
\eea
 There are eight terms which correspond to the eight ways to group the indices $\{r,s,x,y,e,f\}$ into the three factors of inverse propagators, excluding terms of the form $G^{-1}_{ij}$ where $\{i,j\} = \{r,s\}$ or $\{x,y\}$ or $\{e,f\}$. In the  last line of equation (\ref{LAM030}) we use the notation introduced in section \ref{section:RGflow} and write only one term, indicating that there are eight terms in total with the factor (8). 

The functional derivative $\delta G/\delta R$ is calculated in section \ref{section:standardBS} and the result is given in equation (\ref{GderR2}). The method to obtain the second derivative is exactly analogous. Starting from the expression in equation (\ref{GderR2}) and taking an additional derivative we obtain:
\bea
\label{GderR2derR2-a}
-4\frac{\delta}{\delta R_{cd}}\frac{\delta G_{xy}}{\delta R_{ab}} &&= \langle \varphi_c \varphi_d \varphi_a \varphi_b \varphi_x \varphi_y\rangle
-\langle \varphi_c \varphi_d \rangle \langle \varphi_a \varphi_b \varphi_x \varphi_y\rangle \nonumber\\
&& -\langle \varphi_c \varphi_d \varphi_x \varphi_y\rangle\langle \varphi_a \varphi_b \rangle + \langle \varphi_c \varphi_d \rangle \langle \varphi_x \varphi_y \rangle \langle \varphi_a \varphi_b \rangle\nonumber \\[2mm]
&& -\langle \varphi_x \varphi_y \rangle\langle \varphi_c \varphi_d \varphi_a \varphi_b \rangle + \langle \varphi_c \varphi_d \rangle \langle \varphi_x \varphi_y \rangle \langle \varphi_a \varphi_b\rangle\,.
\eea
Converting to connected functions the right side of (\ref{GderR2derR2-a}) becomes:
\bea
\label{GderR2derR2-b}
V^c_{a b c d xy}&&+G_{a y} V^c_{b c d x}+G_{d y} V^c_{a b c x}+G_{c y} V^c_{a b d x}+G_{b y} V^c_{a c d x}+G_{d x} V^c_{a b c y}+G_{c x} V^c_{a b d y}+G_{b x}
   V^c_{a c d y}\nonumber\\
&&+G_{a x} V^c_{b c d y}+G_{b d} V^c_{a c x y}+G_{a d} V^c_{b c x y}+G_{b c} V^c_{a d x y}+G_{a c} V^c_{b d x y}+G_{a y} G_{b d}
   G_{c x}+G_{a y} G_{b c} G_{d x}\nonumber\\
   &&+G_{a d} G_{b y} G_{c x}+G_{a c} G_{b y} G_{d x}+G_{a x} G_{b d} G_{c y}+G_{a d} G_{b x} G_{c
   y}+G_{a x} G_{b c} G_{d y}+G_{a c} G_{b x} G_{d y}\,.
\eea
Combining permutations of external indices to make this result more compact we write:
\bea
\label{GderR2derR2-c}
-4 \frac{\delta^2 G_{xy}}{\delta R_{cd}\delta R_{ab}} &&= V^c_{xyabcd}+ (8) V^c_{xabc}G_{yd}+ (4)V^c_{xyab}G_{cd} +(8)G_{xa}G_{yb}G_{cd} \,.
\eea

Substituting equations (\ref{LAM020}), (\ref{GderR2}), (\ref{LAM030}) and (\ref{GderR2derR2-b}) into (\ref{BS222-a}) produces a lengthy expression that can be manipulated into a compact form. The result is shown in figure \ref{fig:BS222-final}, and some details are given in Appendix \ref{appendix:BS222}. In equilibrium the expression depends only on the differences of co-ordinates $(x_a-x_b)$, $(x_c-x_d)$ and $(x_e-x_f)$, and in momentum space it depends on only 3 momenta. For example, the momentum dependence of the vertex $\Lambda_{abcdef}$ (the second diagram on the right side of figure \ref{fig:BS222-final}) is shown in figure \ref{fig:diag6}. 
\par\begin{figure}[H]
\begin{center}
\includegraphics[width=17cm]{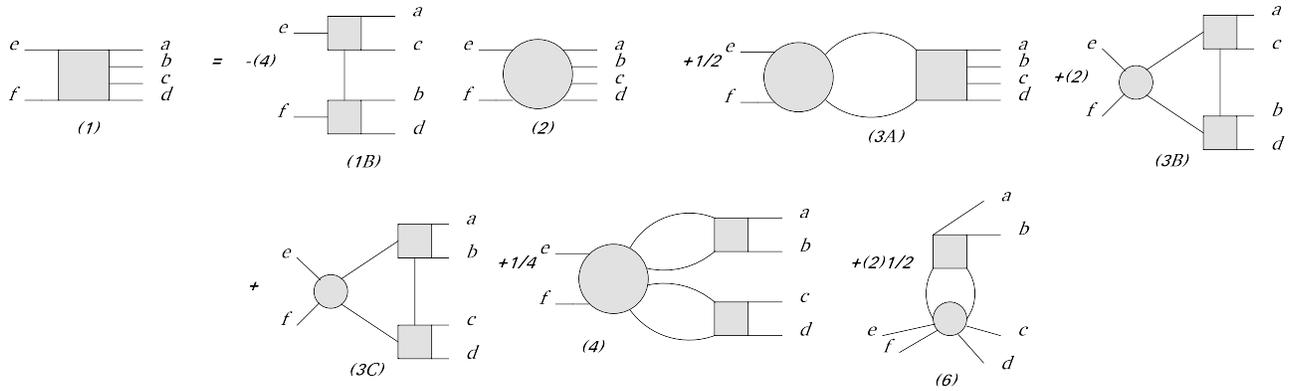}
\end{center}
\caption{The BS equation for 6-point vertex function from the 2PI effective action. The numbers under each graph refer to the discussion in Appendix A. \label{fig:BS222-final}}
\end{figure}
\par\begin{figure}[H]
\begin{center}
\includegraphics[width=3cm]{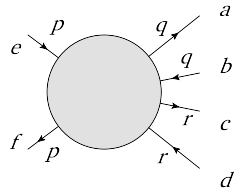}
\end{center}
\caption{Part (2) of figure \ref{fig:BS222-final} with co-ordinate indices and momentum arguments for each leg.\label{fig:diag6}}
\end{figure}

We note that the BS equations in figures \ref{fig:standardBS} and \ref{fig:BS222-final} do not form a closed set, since the diagrams (1B) and (3B) in figure \ref{fig:BS222-final} contain non-diagonal 4-point vertices. In section \ref{section:level2-2PI} we will show that, in spite of this, these BS equations can be used to truncate the RG equations at the level of the second equation, because of the fact that the vertex $\Phi_\kappa^{(4)}$ also contains non-diagonal 4-point vertices. 

\subsection{BS equations from the 4PI effective action}
\label{section:BS-4PI}

In this section we obtain two BS type integral equations from the 4PI effective action. Throughout this paper, in order to avoid a proliferation of indices, we do not introduce additional subscripts to distinguish the 2PI and 4PI effective actions and the vertices obtained from them.

\begin{figure}[H]
\begin{center}
\includegraphics[width=14cm]{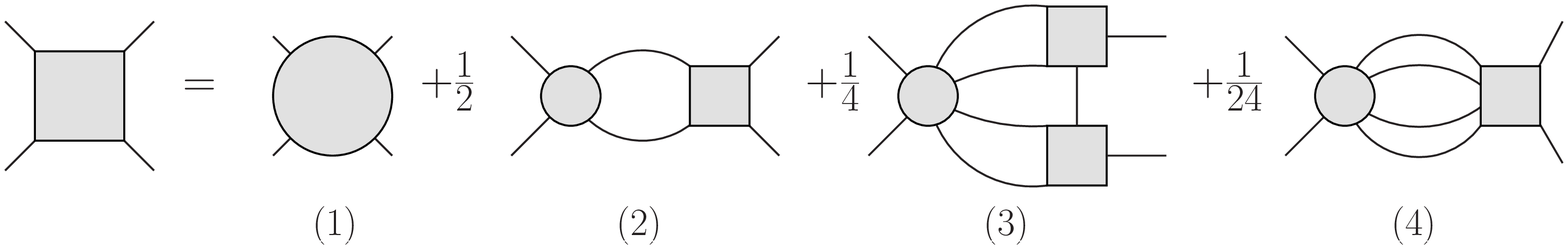}
\end{center}
\caption{The BS equation that is produced by the functional derivative $\delta\Phi/\delta R_{ij}\delta G_{kl}$ where $\Phi$ is the 4PI effective action. Legs on the left side of the figure have indices $(k,l)$ and in momentum space they carry momenta $\pm p$. The legs on the right side have indices $(i,j)$ and carry momenta $\pm q$. \label{fig:BS22-4PI}}
\end{figure}

We can obtain the BS equation for a diagonal 4-point function from the 4PI effective action, following the technique in section \ref{section:standardBS}, by calculating the functional derivative $\delta^3 \Phi/\delta R\delta G$ where $\Phi$ is the 4PI effective action instead of the 2PI one.  In section \ref{section:standardBS} the integral equation for the 4-vertex was obtained by comparing the results of equations (\ref{side1}) and (\ref{side2}). Using the 4PI effective action, equation (\ref{preside1}) contains two additional terms produced by the source $R^{(4)}$, but it is easy to see that the result in equation (\ref{side1}) is unchanged. Equation (\ref{side2}) becomes (in the symmetric theory):
\bea
\label{side2-4PI}
\frac{\delta}{\delta R_{ij}} \frac{\delta}{\delta G_{kl}}\Phi =   \frac{\delta G_{xy}}{\delta R_{ij}}\frac{\delta^2\Phi}{\delta G_{xy}\delta G_{kl}} +  \frac{\delta V_{xywz}}{\delta R_{ij}}\frac{\delta^2\Phi}{\delta V_{xywz}\delta G_{kl}}\,.
\eea
Using equation (\ref{LAM0nl}) the two terms on the right side contain the kernels $\frac{1}{4}(\Lambda^{\rm disco}_{xykl}+\Lambda_{xykl})$ and $\frac{1}{2}\frac{1}{4!}\Lambda_{xywzkl}$ (note that the kernel obtained from $\delta\Phi/\delta G\delta V$ does not have a disconnected piece). The derivative $\delta G/\delta R$ is calculated in section \ref{section:standardBS} and given in equation (\ref{GderR2}).
The derivative $\delta V/\delta R$ can be calculated in exactly the same way. Some details are found in Appendix \ref{appendix:VR}, the result is given in equation (\ref{VderR2}). Substituting the expressions for the derivatives we obtain the result in figure \ref{fig:BS22-4PI}. 
\par

\begin{figure}[H]
\begin{center}
\includegraphics[width=17cm]{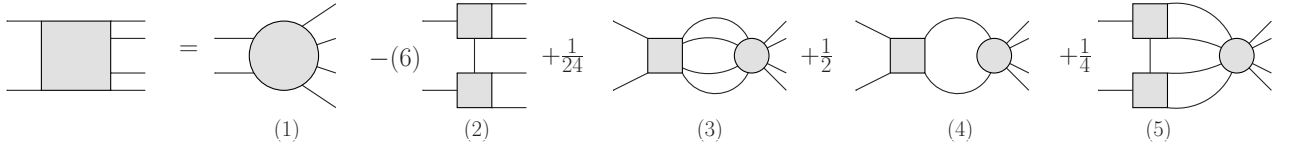}
\end{center}
\caption{The BS equation that is produced by the functional derivative $\delta\Phi/\delta R_{ab}\delta V_{xywz}$. Legs on the left side of the figure have indices $(a,b)$ and in momentum space they carry momenta $\pm p$. The legs on the right side have indices $(x,y,w,z)$ and carry momenta $(k_1,k_2,k_3,-k_1-k_2-k_3)$. \label{fig:BS24}}
\end{figure}

We can also obtain a BS equation for a partially-diagonal 6-point function of the form $V(p,-p,k_1,k_2,k_3,-k_1-k_2-k_3)$ by calculating the functional derivative $\delta\Phi/\delta R\delta V$ where $\Phi$ is the 4PI effective action. Using the chain rule we obtain:
\bea
\label{BS24-a}
0=\frac{\delta^2 \Phi}{\delta R_{ab}\delta V_{cdef}}  = 
\frac{\delta G_{xy}}{\delta R_{ab}}\;\frac{\delta^2\Phi}{\delta G_{xy}\delta V_{cdef}} + \frac{\delta V_{xywz}}{\delta R_{ab}} \; \frac{\delta^2\Phi}{\delta V_{xywz}\delta V_{cdef}}\,,
\eea
where the zero on the left side comes from equation (\ref{side1}). 
The functional derivatives of the effective action give the kernels (see equation (\ref{LAM0nl})):
\bea
&& 2\cdot 4!\,\frac{\delta^2\Phi}{\delta G_{xy}\delta V_{cdef}} = G_{cc^\prime}G_{dd^\prime}G_{ee^\prime}G_{ff^\prime}\big(\Lambda_{xyc^\prime d^\prime e^\prime f^\prime}\big) \,,\nonumber\\
&& 4!\cdot 4!\,\frac{\delta^2\Phi}{\delta V_{xywz}\delta V_{cdef}} = G_{xx^\prime}G_{yy^\prime}G_{ww^\prime}G_{zz^\prime}G_{cc^\prime}G_{dd^\prime}G_{ee^\prime}G_{ff^\prime}\big(\Lambda^{\rm disco}_{x^\prime y^\prime w^\prime z^\prime c^\prime d^\prime e^\prime f^\prime}+\Lambda_{x^\prime y^\prime w^\prime z^\prime c^\prime d^\prime e^\prime f^\prime}\big)\,.
\label{LAM24and44}
\eea
The kernel obtained from $\delta\Phi/\delta G\delta V$ does not contain a disconnected piece. The kernel from $\delta\Phi/\delta V\delta V$ does have a disconnected piece which is produced by the basketball diagram. 
The derivative $\delta G_{xy}/\delta R_{ab}$ is given in equation (\ref{GderR2}) and $\delta V_{xywz}/\delta R_{ab}$ is calculated in Appendix \ref{appendix:VR}. 
To obtain the BS equation we substitute (\ref{GderR2}), (\ref{LAM24and44}) and  (\ref{VderR2}) into (\ref{BS24-a}). After a tedious but straightforward calculation we obtain the result in figure \ref{fig:BS24}.
\par

\section{Higher order truncations}
\label{section:HOtruncations}

\subsection{Truncation at the second level using the 2PI effective action}
\label{section:level2-2PI}

In this section we consider the truncation of the RG flow equations at the level of the second equation using integral equations obtained from the 2PI effective action (figures \ref{fig:standardBS} and \ref{fig:BS222-final}) and following the same procedure as in section \ref{section:LOtruncation}.
The truncation can only be done if we restrict the second RG flow equation (equation (\ref{flow2})) to diagonal external momenta and consider only 4-point functions of the form $\Phi_\kappa^{(4)}(p,-p,q,-q)$, so that the 6-point function that appears in the tadpole diagram has diagonal momentum arguments of the form $\Phi_\kappa^{(6)}(p,-p,q,-q,k,-k)$. The BS equations have kernels with four and six legs of the form $\Lambda_{ijkl}=4\delta\Phi/\delta G_{ij}\delta G_{kl}$ and $\Lambda_{ijklrs}=8\delta\Phi/\delta G_{ij}\delta G_{kl}\delta G_{rs}$ which can be extended to the deformed theory by taking functional derivatives of the effective action  and then evaluating at $G=G_\kappa$. This produces BS 4- and 6-point vertices that depend on the flow parameter.
We show below how to truncate the hierarchy of RG equations at the level of the second equation by replacing the vertices $\Phi_\kappa^{(4)}$ and $\Phi_\kappa^{(6)}$ with the corresponding BS vertices.

\begin{figure}[H]
\begin{center}
\includegraphics[width=17cm]{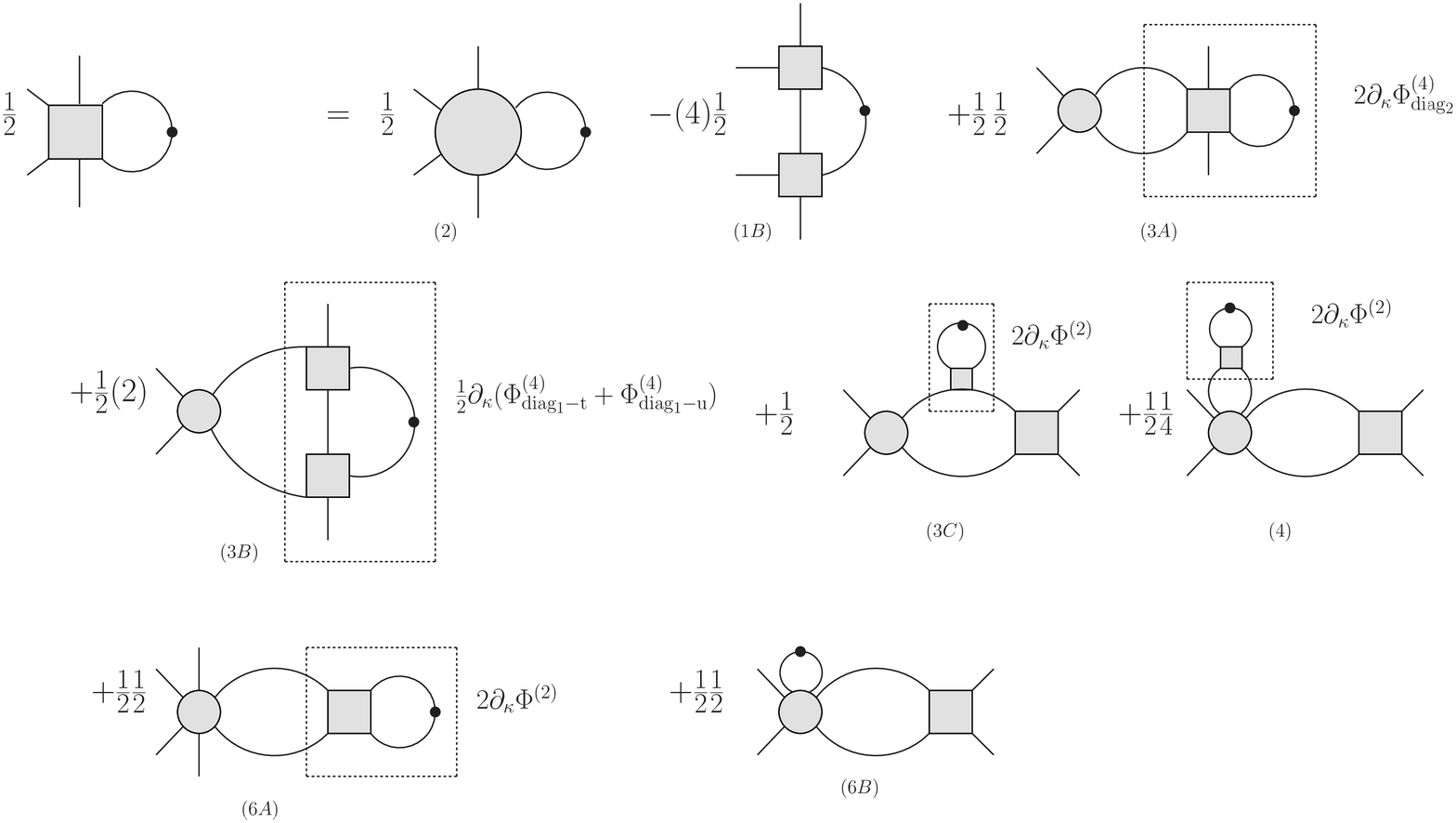}
\end{center}
\caption{The tadpole diagram from the second RG flow equation with diagonal momenta and the 6-point vertex replaced using the BS equation in figure \ref{fig:BS222-final}. In each diagram, the legs on the left side have co-ordinate indices $(e,f)$ and carry momenta $\pm p$. The remaining legs have co-ordinate indices $(c,d)$ and momenta $\pm k$. 
The dotted boxes indicate combinations that can be replaced by re-substituting the RG flow equations. The notation $\partial_\kappa \Phi^{(4)}_{{\rm diag_2}}$ indicates the second diagram on the right side of the  second RG equation in figure \ref{fig:renormGroup}, $\partial_\kappa \Phi^{(4)}_{{\rm diag_1}-t}$ indicates the $t$-channel of the first diagram, etc. The number under each diagram indicates the term in figure \ref{fig:BS222-final} that produced it. The graph labelled (6) in figure \ref{fig:BS222-final} produces two different contributions (6A) and (6B), which correspond respectively to the diagrams in (6) with $(c,d)$ and $(a,b)$ attached to the kernel.  \label{fig:222truncation}}
\end{figure}

\begin{figure}[H]
\begin{center}
\includegraphics[width=17cm]{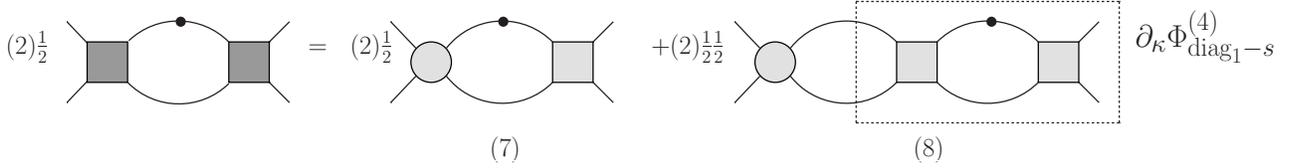}
\end{center}
\caption{The surviving s-channel bubble diagram from the second RG flow equation with the left 4-point vertex replaced using the BS equation in figure \ref{fig:standardBS}. The dotted box on the right side indicates a combination that can be rewritten using the second RG flow equation. \label{fig:222truncationb}}
\end{figure}

\begin{figure}[H]
\begin{center}
\includegraphics[width=17cm]{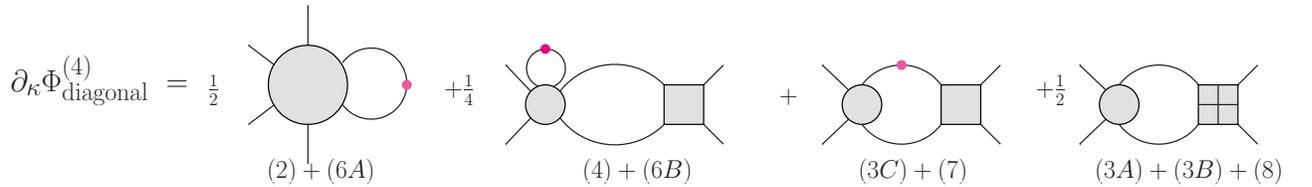}
\end{center}
\caption{The result obtained by combining the surviving diagrams in figures \ref{fig:222truncation} and \ref{fig:222truncationb}. The subscript ``diagonal'' indicates that the momentum arguments are $\pm p$ on the left legs and $\pm k$ on the right legs.  \label{fig:222truncation-all}}
\end{figure}

We start by substituting the BS equation in figure \ref{fig:BS222-final} into the tadpole graph in the second line of figure \ref{fig:renormGroup}. This produces the set of graphs in figure \ref{fig:222truncation}. The diagram labelled (1B) cancels the $t$- and $u$-channels from the bubble graph in the second line of figure \ref{fig:renormGroup}. The remaining $s$-channel diagram can be rewritten by replacing the 4-vertex $\Phi^{(4)}_\kappa$ on the left side with the BS vertex $V_\kappa$ in figure \ref{fig:standardBS} . The result is shown in figure \ref{fig:222truncationb}.  Combining the results in figures \ref{fig:222truncation} and \ref{fig:222truncationb} gives the result in figure \ref{fig:222truncation-all}, where we have used the notation in figure \ref{fig2:Gkappa-deriv} to represent the sum of terms $\partial_\kappa R_\kappa$ (represented by a black dot) and $\partial_\kappa \Phi^{(2)}$ by a small grey dot (red on-line). 
\par
\par
\par

We now show that this truncation has the same property as the lower order truncation, that the flow is a total derivative with respect to the flow parameter (see equation (\ref{2point-kappa})). 
Differentiating the BS 4-vertex in equation (\ref{BSfirst-mom-ext}) we obtain:
\bea
\label{22der}
\partial_\kappa (V_{efcd})_\kappa && = \partial_\kappa (\Lambda_{efcd})_\kappa+\frac{1}{2}\, \partial_\kappa(\Lambda_{efxy})_\kappa\;(G_{xz})_\kappa (G_{yw})_\kappa (V_{zwcd})_{\kappa}
+ (\Lambda_{efxy})_\kappa\,\partial_\kappa(G_{xz})_\kappa (G_{yw})_\kappa \;(V_{zwcd})_{\kappa}\nonumber\\
&&+\frac{1}{2}\, (\Lambda_{efxy})_\kappa\;(G_{xz})_\kappa (G_{yw})_\kappa\; \partial_\kappa(V_{zwcd})_{\kappa}\,.
\eea
The third term in this equation is the third diagram on the right side of figure \ref{fig:222truncation-all}, and the fourth term is the fourth diagram if we identify $V_\kappa = \Phi^{(4)}_\kappa$. In order to obtain a diagrammatic representation of the first and second terms we need to calculate the derivative of the kernel. Using the fact that all $\kappa$  dependence enters through the propagator we have:
\bea
\label{LAM020der}
\partial_\kappa(\Lambda_{efxy})_\kappa = 4\partial_\kappa\;\frac{\delta^2\Phi}{\delta G_{ef}\delta G_{xy}}\bigg| = 
4\partial_\kappa (G_{cd})_\kappa \frac{\delta^3\Phi}{\delta G_{cd}\delta G_{ef}\delta G_{xy}}\bigg|  = \frac{1}{2}\partial_\kappa (G_{cd})_\kappa (\Lambda_{cdefxy})_\kappa \,. \eea
This result is the first diagram in figure \ref{fig:222truncation-all}, and substituting (\ref{LAM020der}) into the second term of (\ref{22der}) gives the second diagram.  Combining all pieces, we obtain:
\bea
\label{222result}
\partial_\kappa \Phi_\kappa^{(4)}(p,-p,q,-q) = \partial_\kappa V_\kappa(p,-p,q,-q)\,,
\eea
which shows that the truncation has the property that the flow is a total derivative.

\subsection{Truncation at the second level using the 4PI effective theory }

The result of the previous section shows that the RG flow equations are equivalent to integral equations which can be obtained from the 2PI effective action, if one considers only the vertex $\Phi_\kappa^{(4)}$ with diagonal momentum arguments. 
Since the RG equations are obtained by introducing a bi-local source term, one might suspect that the correspondence between the flow equations and $n$PI integral equations is limited to the special case of diagonal vertices and the 2PI effective action.
%The second RG equation for the non-diagonal vertex $\Phi_\kappa^{(4)}(p_1,p_2,p_3,-p_1-p_2-p_3)$ contains a partially-diagonal 6-point vertex of the form $\Phi\kappa^{(6)}(p,-p,q_1,q_2,q_3,-q_1-q_2-q_3)$.
In this section we show that the correspondence holds at higher orders. 
%To see if the correspondence holds for non-diagonal $n$-point functions, we can use the BS equation for the vertex $V(p,-p,q_1,q_2,q_3,-q_1-q_2-q_3)$ obtained in section \ref{section:BS-4PI} to truncate the second RG equation. 
We truncate the RG equations using the BS equations in figures \ref{fig:BS22-4PI} and \ref{fig:BS24} extended to the deformed theory as in the previous section. 

\begin{figure}[H]
\begin{center}
\includegraphics[width=16cm]{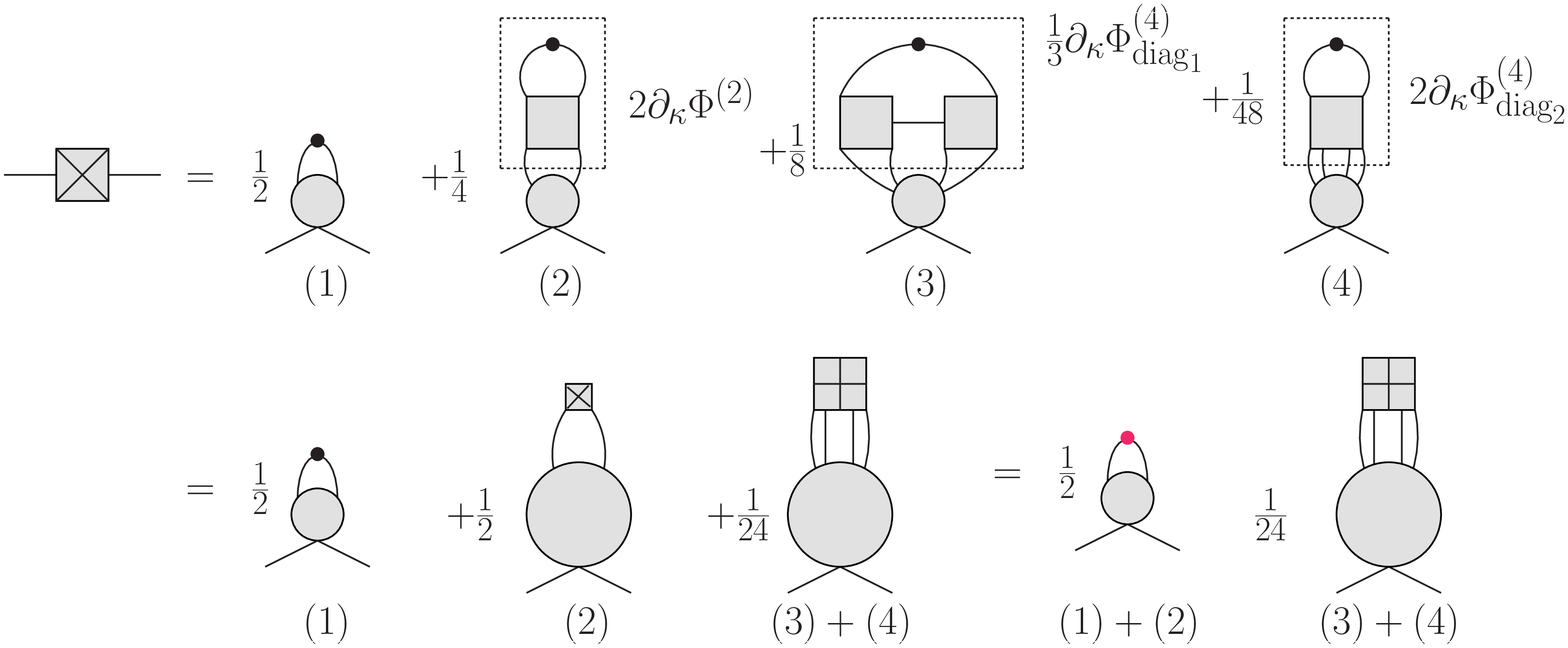}
\end{center}
\caption{The diagrams produced when the BS equation in figure \ref{fig:BS22-4PI} is substituted into the tadpole diagram in the first line of figure \ref{fig:renormGroup}. In the second line we regroup the graphs by re-substituting the RG equations and using figure \ref{fig2:Gkappa-deriv}. The numbers under each diagram indicate the pieces of figure \ref{fig:BS22-4PI} that contributed. \label{fig:22truncation-4PI}}
\end{figure}

\begin{figure}[H]
\begin{center}
\includegraphics[width=17cm]{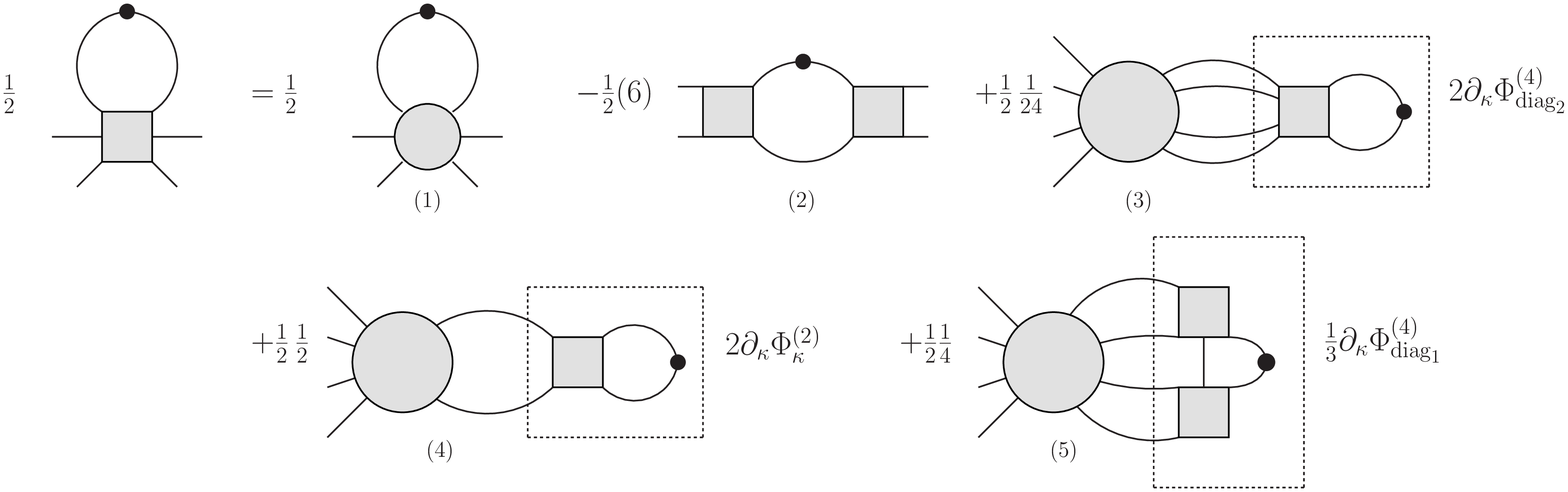}
\end{center}
\caption{The result of substituting the 6-point vertex in figure \ref{fig:BS24} into the tadpole diagram in the second line of figure \ref{fig:renormGroup}.  \label{fig:BS24truncation}}
\end{figure}

\begin{figure}[H]
\begin{center}
\includegraphics[width=12cm]{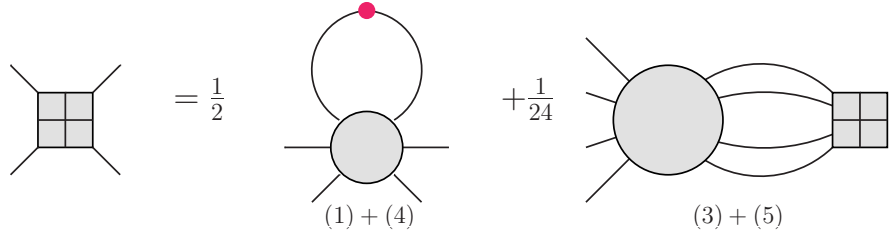}
\end{center}
\caption{The second RG equation truncated with the BS equation in figure \ref{fig:BS24}. The numbers under the diagrams correspond to the pieces of figure \ref{fig:BS24truncation} that were combined. The diagram labelled (2) in figure \ref{fig:BS24truncation} cancels the bubble graph in figure \ref{fig:renormGroup}. \label{fig:BS24truncation-all}}
\end{figure}

Substituting figure \ref{fig:BS22-4PI} into the tadpole diagram in the first line of figure \ref{fig:renormGroup} produces the diagrams in figure \ref{fig:22truncation-4PI}. Similarly, using figure \ref{fig:BS24} the tadpole diagram the second equation in figure \ref{fig:renormGroup} takes the form shown in figure \ref{fig:BS24truncation}. Substituting figure \ref{fig:BS24truncation} into the second line in figure \ref{fig:renormGroup} we obtain the result in figure \ref{fig:BS24truncation-all}.
\par
\par
\par

Now we compare our results for $\partial_\kappa\Phi_\kappa^{(2)}$ and $\partial_\kappa\Phi_\kappa^{(4)}$ in figures \ref{fig:22truncation-4PI} and \ref{fig:BS24truncation-all} with the derivatives $\partial_\kappa\Sigma_\kappa$ and $\partial_\kappa V_\kappa$  of the 2- and 4-point functions obtained from the 4PI equations of motion and extended to the deformed theory. Note that the vertex $V_\kappa$ depends on the flow parameter $\kappa$ only indirectly through the fact that the equation of motion for the 4-point function is coupled to the equation of motion for the 2-point function, which is evaluated at $G=G_\kappa$ after the functional derivatives are taken.

In the 4PI theory equation (\ref{sigma-2PI-b}) becomes:
\bea
\label{sigma-4PI-b}
\partial_\kappa(\Sigma_{ij})_\kappa &&= 2\partial_\kappa (G_{kl})_\kappa \frac{\delta^2\Phi}{\delta G_{kl}\delta G_{ij}}\bigg|_{G_\kappa} \nonumber\\
&&+
2\partial_\kappa (V_{klrs})_\kappa (G_{ka})_\kappa (G_{lb})_\kappa  (G_{rc})_\kappa  (G_{sd})_\kappa  \;G^{-1}_{aa^\prime}G^{-1}_{bb^\prime}G^{-1}_{cc^\prime}G^{-1}_{dd^\prime}  \frac{\delta^2\Phi}{\delta V_{a^\prime b^\prime c^\prime d^\prime}\delta G_{ij}}\bigg|_{G_\kappa} \nonumber\\
&& = \frac{1}{2}\partial_\kappa (G_{kl})_\kappa (\Lambda_{ijkl})_\kappa + \frac{1}{24}\,\partial_\kappa (V_{klrs})_\kappa (G_{ka})_\kappa (G_{lb})_\kappa  (G_{rc})_\kappa  (G_{sd})_\kappa  (\Lambda_\kappa)_{abcdij}\,.
\eea
Equation (\ref{sigma-4PI-b}) is precisely the result that is shown in the last line of figure \ref{fig:22truncation-4PI} if we identify $V_\kappa = \Phi^{(4)}_\kappa$, which means we have obtained the equation 
\bea \label{2result}
\partial_\kappa \Phi_\kappa^{(2)}(p) = \partial_\kappa \Sigma_\kappa(p)\,,
\eea
 at the level of the 4PI effective action. 

For the 4-point function we have:
\bea
\partial_\kappa (V_{xywz})_{\kappa} && =4!\partial_\kappa \bigg[G^{-1}_{xx^\prime}G^{-1}_{yy^\prime}G^{-1}_{ww^\prime}G^{-1}_{zz^\prime}  \frac{\partial^2\hat\Phi}{\delta V_{x^\prime y^\prime w^\prime z^\prime }}\bigg]_{G=G_\kappa}\nonumber\\
&& =4!\partial_\kappa (G_{ab})_\kappa \bigg[ \frac{\delta}{\delta G_{ab}}\bigg( G^{-1}_{xx^\prime}G^{-1}_{yy^\prime}G^{-1}_{ww^\prime}G^{-1}_{zz^\prime}  \;\frac{\partial\hat\Phi}{\delta V_{x^\prime y^\prime w^\prime z^\prime }}\bigg)\bigg]_{G=G_\kappa} \nonumber\\
&&~~~~~~ + \partial_\kappa (V_{abcd})_\kappa \bigg[ \frac{\delta}{\delta V_{abcd}}\bigg(G^{-1}_{xx^\prime}G^{-1}_{yy^\prime}G^{-1}_{ww^\prime}G^{-1}_{zz^\prime} \;\frac{\partial\hat\Phi}{\delta V_{xywz}}\bigg)\bigg]_{G=G_\kappa}\nonumber\\
&&  =\frac{1}{2} \partial_\kappa (G_{ab})_\kappa (\Lambda_{abxywz})_\kappa + \frac{1}{4!} \partial_\kappa (V_{abcd})_\kappa (G_{ax})_\kappa (G_{by})_\kappa(G_{cw})_\kappa(G_{dz})_\kappa(\Lambda_{abcdxywz})_\kappa\,.
\label{Vder-kappa}
 \eea
The hat indicates that the basketball diagram has been removed from the effective action (since this is the diagram that produces the left side in the V equation of motion in figure \ref{fig:Veom} - see the discussion under equation (\ref{Veom})). The basketball diagram does not contribute to either the 6-point or 8-point kernel (it produces $\Lambda_{abcdxywz}^{\rm disco}$). There are no contributions from the functional derivative with respect to $G$ acting on the inverse propagators because the sum of these terms gives zero using the $V$ equation of motion.
Equation (\ref{Vder-kappa}) is exactly the equation shown in figure \ref{fig:BS24truncation-all} if we use $ V_\kappa =  \Phi^{(4)}_\kappa$. Thus we have obtained:
\bea
\label{4result}
\partial_\kappa \Phi_\kappa^{(4)}(p_1,p_2,p_3,-p_1-p_2-p_3) = \partial_\kappa V_\kappa(p_1,p_2,p_3,-p_1-p_2-p_3)\,,
\eea
which is the generalization of equation (\ref{222result}) to non-diagonal momenta.

Equations (\ref{2result}) and (\ref{4result}) show that the RG equations for the 2- and 4-point functions are total derivatives whose integrals can be written as the  4PI equations of motion. 
 
\section{Conclusions}
\label{section:conclusions}

In this paper we have studied the connection between two different formalisms that are commonly used to study non-perturbative systems: the exact renormalization group and $n$-particle irreducible  effective theories. 
We have derived Bethe-Salpeter type integral equations that can be used to truncate the RG flow equations, and shown that the resulting flow equations for the 2- and 4-point functions are total derivatives whose integrals give the $n$PI eom's. Since the full hierarchy of  RG flow equations are obtained using a single bi-local source term, this result is surprising and suggests that a BS truncation at arbitrary orders produces equations whose integrals gives the $n$PI eom's. This establishes a direct connection between two non-perturbative methods. 
It also means that the truncation of the RG equations at any level of the hierarchy can be systematically extended by adding more and more skeleton diagrams to the effective action. For the $n$PI formalism, there might be a practical advantage in reformulating the integral equations as flow equations, because initial value problems are usually easier to solve than non-linear integral equations.

\section*{Acknowledgements}

This work was supported by the Natural and Sciences and Engineering Research Council of Canada.

\appendix

\section{Derivation of the BS equation for the diagonal 6-point vertex from the 2PI effective action}
\label{appendix:BS222}

\begin{figure}[H]
\begin{center}
\includegraphics[width=17cm]{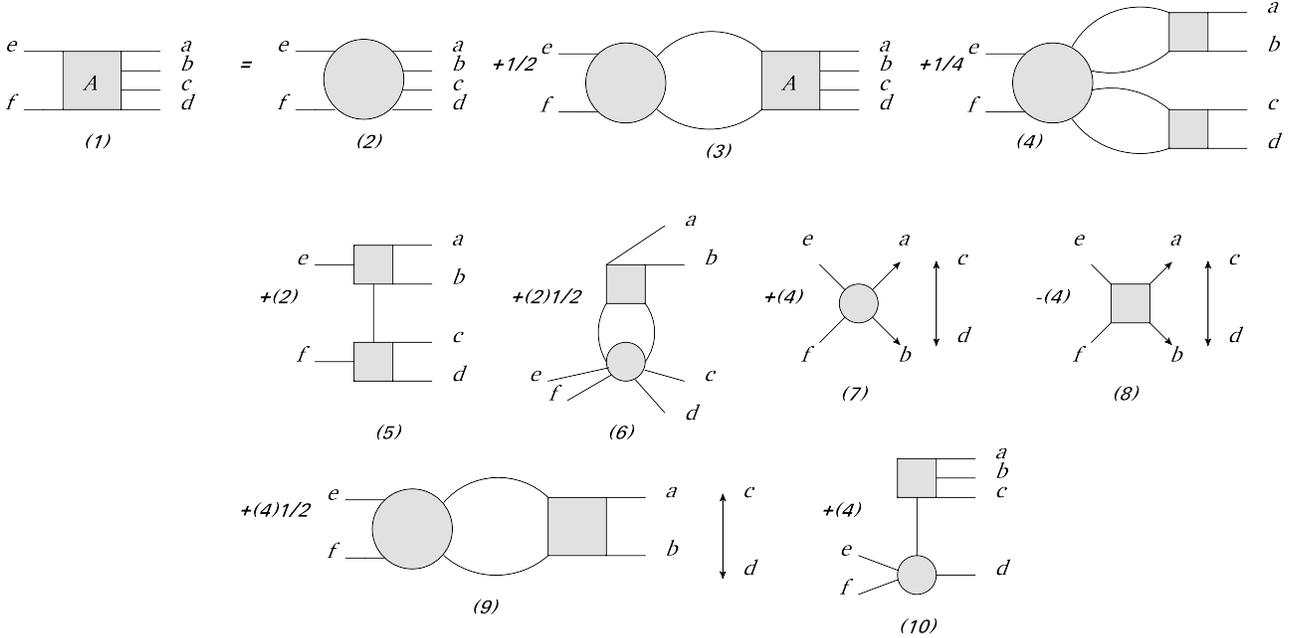}
\end{center}
\caption{The diagonal 6-point vertex function in terms of amputated vertices.  \label{fig:BS222-part1}}
\end{figure}

In this appendix we give some details of the derivation of the result in figure \ref{fig:BS222-final}.
Substituting equations (\ref{LAM020}), (\ref{GderR2}), (\ref{LAM030}) and (\ref{GderR2derR2-b}) into (\ref{BS222-a}) one finds immediately that all terms that contain a factor $\delta_{ij}$ cancel exactly. There are 10 remaining terms, which are shown in figure \ref{fig:BS222-part1}. The 6-point boxes in this figure are amputated vertex functions, which is indicated by the letter `$A$' inside each box. 
\par
\begin{figure}[H]
\begin{center}
\includegraphics[width=16cm]{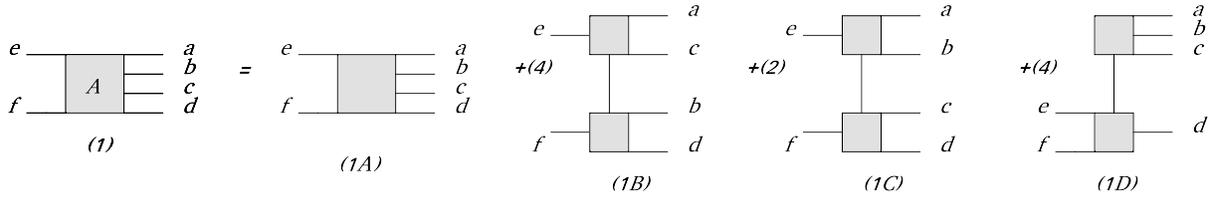}
\end{center}
\caption{The diagram in part (1) of figure \ref{fig:BS222-part1} in terms of proper vertices.  \label{fig:expDiag1}}
\end{figure}
\begin{figure}[H]
\begin{center}
\includegraphics[width=14cm]{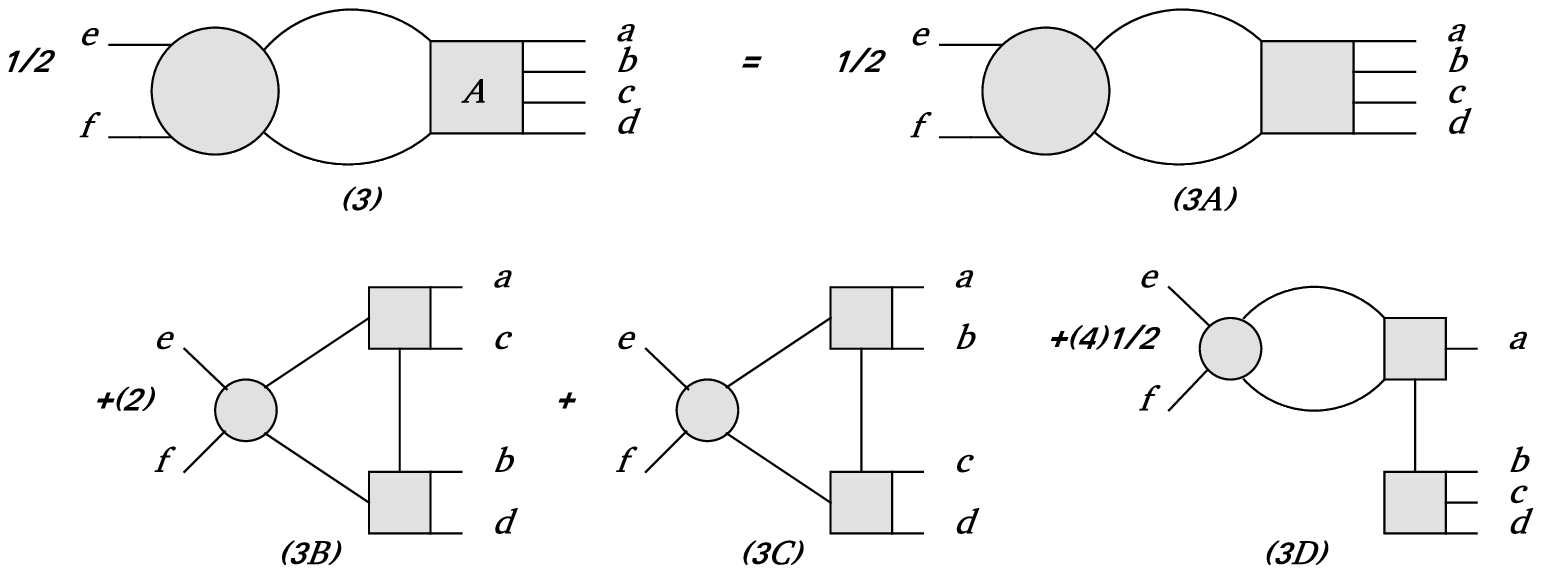}
\end{center}
\caption{The diagram in part (3) of figure \ref{fig:BS222-part1} in terms of proper vertices.  \label{fig:expDiag3}}
\end{figure}

\noindent The diagrams in parts (7), (8) and (9) of the figure contain disconnected contributions, but one immediately sees that they cancel using the lower-order BS equation in equation (\ref{BSfirst-coord}). Parts (1) and (3) of the figure can be written in terms of proper vertex functions using:
\bea
V^{\rm amputated}_{a b c d e f} &&=V_{a b c d e f}+  G_{z_7 z_8} V_{a b c z_8} V_{d e f z_7}+G_{z_7 z_8} V_{a b d z_8} V_{c e f z_7}+G_{z_7 z_8} V_{a c d z_8} V_{b e f z_7}+G_{z_7
   z_8} V_{a e f z_7} V_{b c d z_8}\nonumber\\[2mm]
   &&+G_{z_7 z_8} V_{a b e z_8} V_{c d f z_7}+G_{z_7 z_8} V_{a c e z_8} V_{b d f z_7}+G_{z_7 z_8}
   V_{a d f z_7} V_{b c e z_8}+G_{z_7 z_8} V_{a d e z_8} V_{b c f z_7}\nonumber\\[2mm]
   &&+G_{z_7 z_8} V_{a c f z_7} V_{b d e z_8}+G_{z_7 z_8} V_{a b
   f z_7} V_{c d e z_8}\,, \nonumber\\[2mm]
   &&= V_{a b c d e f}+(10)G_{z_7 z_8} V_{a b c z_8} V_{d e f z_7}\,.
\eea
The results are shown in figures \ref{fig:expDiag1} and \ref{fig:expDiag3}.  One can see immediately that (1C) cancels (5), and (1D) and (3D) cancel (10) when we use the BS equation for the 4-vertex in (\ref{BSfirst-coord}) in the lower vertex in (10). The survivors are shown in figure \ref{fig:BS222-final}.

\par
\par

\section{Derivation of the functional derivative $\delta V/\delta R$}
\label{appendix:VR}

In this appendix we calculate $\delta V_{xywz}/\delta R_{ab}$ using the same technique as in sections \ref{section:standardBS} and \ref{section:6diag}. In the symmetric theory we have:
\bea
\label{VderR2-a}
\frac{\delta}{\delta R_{ab}} V_{xywz}=\frac{\delta}{\delta R_{ab}}\bigg(G^{-1}_{xx^\prime}G^{-1}_{xx^\prime}G^{-1}_{xx^\prime}G^{-1}_{xx^\prime}V^c_{x^\prime y^\prime w^\prime z^\prime}\bigg)\,.
\eea
We separate the contributions from the derivative acting on the inverse propagators and the connected vertex function. If the derivative acts only on the vertex function we have:
\bea
\label{VderR2part1}
~~&&G^{-1}_{xx^\prime}G^{-1}_{yy^\prime}G^{-1}_{ww^\prime}G^{-1}_{zz^\prime}\frac{\delta}{\delta R_{ab}} V^c_{x^\prime y^\prime w^\prime z^\prime} \nonumber\\
&& = G^{-1}_{xx^\prime}G^{-1}_{yy^\prime}G^{-1}_{ww^\prime}G^{-1}_{zz^\prime} \frac{\delta}{\delta R_{ab}}\big( \langle \varphi_{x^\prime} \varphi_{y^\prime} \varphi_{w^\prime} \varphi_{z^\prime}\rangle - \langle\varphi_{x^\prime}\varphi_{y^\prime} \rangle \langle\varphi_{w^\prime}\varphi_{z^\prime} \rangle
- \langle\varphi_{x^\prime}\varphi_{w^\prime} \rangle \langle\varphi_{y^\prime}\varphi_{z^\prime} \rangle
- \langle\varphi_{x^\prime}\varphi_{z^\prime} \rangle \langle\varphi_{y^\prime}\varphi_{w^\prime} \rangle \big)\,, \nonumber\\
&& =\frac{i}{2} G^{-1}_{xx^\prime}G^{-1}_{yy^\prime}G^{-1}_{ww^\prime}G^{-1}_{zz^\prime} \bigg(\langle \varphi_{x^\prime} \varphi_{y^\prime} \varphi_{w^\prime} \varphi_{z^\prime} \varphi_a \varphi_b\rangle
- G_{ab} \langle \varphi_{x^\prime} \varphi_{y^\prime} \varphi_{w^\prime} \varphi_{z^\prime}\rangle
-(6)G_{x^\prime y^\prime} \langle \varphi_a \varphi_b \varphi_{w^\prime} \varphi_{z^\prime}\rangle\ \nonumber\\
&&~~~~~~~~~~~~~~~~~~~+2(3) G_{a b} G_{x^\prime y^\prime} G_{wz}\bigg)\,,\nonumber\\
&& = \frac{i}{2}\bigg((8) G_{a z_1} \delta _{b z} V_{w x y z_1}+(4) G_{z_3 z_4} G_{a z_1} G_{b z_2} V_{x z_1 z_2 z_3} V_{w y z z_4}+ (6)   G_{z_3 z_4} G_{a z_1} G_{b z_2} V_{x z z_2 z_3} V_{w y z_1 z_4}\nonumber\\
&&~~~~~~+G_{a z_1} G_{b z_2} V_{w x y z z_1 z_2}\bigg)\,.
   \eea
   Now we consider the contribution obtained when the derivative acts only on the inverse propagators. 
To differentiate the inverse propagators we use:
\bea
\label{chainX}
\frac{\delta G^{-1}_{ij}}{\delta R_{ab}} && = \frac{\delta G^{-1}_{ij}}{\delta G_{mn}}\frac{\delta G_{mn}}{\delta R_{ab}}\,,
\eea
with the result for $\delta G_{mn}/\delta R_{ab}$ given in equation (\ref{GderR2}) and the derivative of the inverse propagator given by:
\bea
\label{gert}
-2\frac{\delta G^{-1}_{ij}}{\delta G_{mn}} = G^{-1}_{im} G^{-1}_{jn} + G^{-1}_{in} G^{-1}_{jm}\,.
\eea  
Using equations (\ref{GderR2}), (\ref{chainX}) and (\ref{gert}) we obtain:
\bea
\label{VderR2part2}
V^c_{x^\prime y^\prime w^\prime z^\prime}\frac{\delta}{\delta R_{ab}}\bigg(G^{-1}_{xx^\prime}G^{-1}_{xx^\prime}G^{-1}_{xx^\prime}G^{-1}_{xx^\prime}\bigg) = -\frac{i}{2}\bigg((8) G_{a z_1} \delta _{b z} V_{w x y z_1}+(4) G_{z_3 z_4} G_{a z_1} G_{b z_2} V_{x z_1 z_2 z_3} V_{w y z z_4}  \bigg)\,. \nonumber\\ \,.
   \eea
Substituting the results in equations (\ref{VderR2part1}) and (\ref{VderR2part2}) into (\ref{VderR2-a}) we obtain:
\bea 
\label{VderR2}
\frac{\delta}{\delta R_{ab}} V_{xywz} = \frac{i}{2}G_{a z_1} G_{b z_2}\bigg(V_{w x y z z_1 z_2} + (6) G_{z_3 z_4}  V_{x z z_2 z_3} V_{w y z_1 z_4}\bigg)\,.
\eea

\end{document}